\documentclass[12pt]{article}
\usepackage{graphicx}

\begin{document}

\title{Computing collinear 4-Body Problem central configurations with given masses}

\author{E. Pi\~na\thanks{On sabbatic leave from Universidad Aut\'onoma Metropolitana} \ Professor {\sl "Eugenio M\'endez Docurro"}\\
de la Escuela Superior de F\'\i sica y Matem\'aticas del IPN \\
Zacatenco, M\'exico, D. F. \\
e-mail: pge@xanum.uam.mx}
\date{}
\maketitle

\abstract{An interesting description of a collinear configuration of four particles is found in terms of two spherical coordinates. An algorithm to compute the four coordinates of particles of a collinear Four-Body central configuration is presented by using an orthocentric tetrahedron, which edge lengths are function of given masses. Each mass is placed at the corresponding vertex of the tetrahedron. The center of mass (and orthocenter) of the tetrahedron is at the origin of coordinates. The initial position of the tetrahedron is placed with two pairs of vertices each in a coordinate plan, the lines joining any pair of them parallel to a coordinate axis, the center of masses of each and the center of mass of the four on one coordinate axis. From this original position the tetrahedron is rotated by two angles around the center of mass until the direction of configuration coincides with one axis of coordinates. The four coordinates of the vertices of the tetrahedron along this direction determine the central configuration by finding the two angles corresponding to it. The twelve possible configurations predicted by Moulton's theorem are computed for a particular mass choice.
}
\

{\sl Keywords:} Four-Body Problem. Collinear central configuration.

\

PACS 45.50.Pk Celestial mechanics 95.10.Ce Celestial mechanics(including n-body problems)

\newpage

\section{Introduction}

The main purpose of this paper is the efficient computation of the collinear central configurations corresponding to a given choice of four masses. A general introduction to the subject of central configurations is the Saari's book \cite{sa}. The collinear four body central configurations is a particular case of the general collinear N-Body central configurations studied by Moulton \cite{mu}. He determines one half of $N!$ classes of different collinear configurations of $N$ particles, and a formal construction of each by a long-time consuming iterated algorithm. The only three-dimensional Four-Body central configuration is an equilateral tetrahedron found by Lehmann-Filh\'es \cite{lf}. An algorithm to compute the non-collinear Four-Body planar central configurations was distributed in the web \cite{ar}. The collinear Four-Body central configurations are now considered in this paper using again the coordinate system published in reference  \cite{pi}. This set of coordinates seem to be very useful for the study of these problems.

We have computed previously central configurations of the Four-Body Problem for the three-dimensional and the planar non-collinear cases in \cite{pi}, \cite{pl}, and \cite{ar}. It is natural to consider at present the collinear case that was previously ignored. The new four-body coordinates is again the tool to attain the task \cite{pi}. That coordinate system for collinear problems is so much simplified that the first action is to present that reduction in the number of coordinates.

The masses of the four particles $m_1$, $m_2$, $m_3$ and $m_4$ are positive, generally different, but the values could be repeated. An Euclidean inertial system of coordinates is the frame to determine the position coordinates of the four particles with the center of mass of them at the origin of coordinates. Then with no loss of generality the four particle positions are function of nine generalized coordinates, which for the collinear case reduce to only four coordinates. Starting from the general case, the Euclidean coordinate system is transformed to the frame of principal axes of inertia \cite{ll} by means of a three dimensional rotation $\bf G$ parameterized by three independent coordinates. Three more coordinates are introduced, as scale factors $R_1$, $R_2$, $R_3$, which are three directed distances closely related to the three principal inertia moments through
\begin{equation}
I_1 = \mu (R_2^2 + R_3^2)\, , \quad I_2 = \mu (R_3^2 + R_1^2) \, , \quad \mbox{and} \quad I_3 = \mu (R_1^2 + R_2^2) \, ,
\end{equation}
where $\mu$ is the mass
\begin{equation}
\mu = \sqrt[3]{\frac{m_1 \, m_2\, m_3\, m_4}{m_1 + m_2 + m_3 + m_4}} \, .
\end{equation}

With the first rotation and the change of scale the resulting four-body configuration has a moment of inertia tensor with the three principal moments of inertia equal. A second rotation $\bf G'$ does not change this property.

The cartesian coordinates of the four particles, with the center of gravity at the origin, written in terms of the new coordinates are
\begin{equation}
\left(\begin{array}{cccc}
x_1 & x_2 & x_3 & x_4 \\
y_1 & y_2 & y_3 & y_4 \\
z_1 & z_2 & z_3 & z_4
\end{array} \right) = {\bf G} \left(\begin{array}{ccc}
R_1 & 0 & 0 \\
0 & R_2 & 0 \\
0 & 0 & R_3
\end{array} \right) {\bf G'}^{\rm T}
\left(\begin{array}{cccc}
a_1 & a_2 & a_3 & a_4 \\
b_1 & b_2 & b_3 & b_4 \\
c_1 & c_2 & c_3 & c_4
\end{array} \right)\, ,
\end{equation}
where $\bf G$ and $\bf G'$ are two rotation matrices, each one a function of three independent coordinates such as the Euler angles, and where the column elements of the constant matrix
\begin{equation}
{\bf E} = \left(\begin{array}{cccc}
a_1 & a_2 & a_3 & a_4 \\
b_1 & b_2 & b_3 & b_4 \\
c_1 & c_2 & c_3 & c_4
\end{array} \right)\, ,
\end{equation}
are the coordinates of the four vertices of a rigid orthocentric tetrahedron, with the center of mass at the origin of coordinates, namely:
\begin{equation} 
\begin{array}{c}
a_1 m_1 + a_2 m_2 + a_3 m_3 + a_4 m_4 = 0 \, , \\
b_1 m_1 + b_2 m_2 + b_3 m_3 + b_4 m_4 = 0 \, , \\
c_1 m_1 + c_2 m_2 + c_3 m_3 + c_4 m_4 = 0 \, .
\end{array}\, . \label{base}
\end{equation}

We used the following notation for the matrix
\begin{equation}
{\bf M} = \left( \begin{array}{cccc}
m_1 & 0 & 0 & 0 \\
0 & m_2 & 0 & 0 \\
0 & 0 & m_3 & 0 \\
0 & 0 & 0 & m_4
\end{array} \right) \, .
\end{equation}

An equivalent condition in order to have three equal inertia moments for the rigid tetrahedron is expressed as
\begin{equation} 
{\bf E\, M\, E}^{\rm T} = \mu \left( \begin{array}{ccc}
1 & 0 & 0 \\
0 & 1 & 0 \\
0 & 0 & 1
\end{array} \right) \label{ort}
\end{equation}

An orthocentric tetrahedron has the property that the perpendicular lines to the faces trough the four vertices intersect at the same point. Orthocentric tetrahedra were considered by Lagrange in 1773 \cite{la}. Other old references on orthocentric tetrahedra are found in a paper by Court \cite{co}, he calls them orthocentric and orthogonal. Placing the four masses at the corresponding vertices, the intersection point is actually the center of mass of the four masses, and the moment of inertia tensor of the four particles has the same principal value in any direction. Equation (\ref{ort}) imply that the inertia tensor of the rigid tetrahedron is proportional by a factor $2 \mu$ to the unit matrix.

The proof of these properties is found in reference \cite{pi}, where we use the notation $m = m_1 + m_2 + m_3 + m_4$ for the total mass of the system.
It was shown that the position vector of one vertex is orthogonal to the three vectors between two vertices of the opposite face (to the position vector of the vertex.)  The square of the distance between two vertices is given by
\begin{equation}
(a_i - a_j)^2 + (b_i - b_j)^2 + (c_i - c_j)^2 = \mu \left( \frac{1}{m_i} + \frac{1}{m_j}\right)\, . \label{lados}
\end{equation}
This last is the condition to have a moment of inertia tensor with the same three principal moments of inertia. The six edges of the tetrahedron should be equal (for an arbitrary $\mu$) to the square root of the right-hand side of this equation. The volume of this tetrahedron is equal to 1/6 if $\mu$ is selected as above.

A different important coordinate system to fix the origin for measuring the $\bf G'$ rotation is presented in this preprint. Two other coordinate systems were presented in references \cite{ar} and \cite{pi}. In this paper the original position of the orthocentric tetrahedron is chosen with two pairs of particles each in a coordinate plan, the line joining two of them orthogonal to a coordinate axis, the center of masses of each and the center of mass of the four coinciding with one coordinate axis. More explicitly particle 1 has the coordinates
\begin{equation}
(a_1, b_1, c_1) = \left(0, \sqrt{\frac{\mu (m_3 + m_4)}{m (m_1 + m_2)}}, \sqrt{\frac{\mu m_2}{m_1 (m_1 + m_2)}}\right)\, .
\end{equation}
Particle 2 has the coordinates
\begin{equation}
(a_2, b_2, c_2) = \left(0, \sqrt{\frac{\mu (m_3 + m_4)}{m (m_1 + m_2)}}, - \sqrt{\frac{\mu m_1}{m_2 (m_1 + m_2)}}\right)\, .
\end{equation}
Particle 3 has the coordinates
\begin{equation}
(a_3, b_3, c_3) = \left(\sqrt{\frac{\mu m_4}{m_3 (m_3 + m_4)}}, - \sqrt{\frac{\mu (m_1 + m_2)}{m (m_3 + m_4)}}, 0\right)\, .
\end{equation}
Particle 4 has the coordinates
\begin{equation}
(a_4, b_4, c_4) = \left(- \sqrt{\frac{\mu m_3}{m_4 (m_3 + m_4)}}, - \sqrt{\frac{\mu (m_1 + m_2)}{m (m_3 + m_4)}}, 0\right)\, .
\end{equation}

This rigid tetrahedron is the generalization of the rigid triangle of the Three-Body problem with the center of mass at the orthocenter discussed previously in \cite{pb}. The same triangle was used with different purposes by C. Sim\'o \cite{si}.

Assuming the four particles move on a straight line, Moulton \cite{mu} finds the motion of the line is in a constant plane. The number of Euclidean coordinates changes from twelve to eight. The first rotation around the center of mass is a function of one angle $\psi$ that rotate the line from its position in the Euclidean plane to the one dimensional line where the relative positions move. The three scales $R_i$ are reduced to one scale $R_1 = R$. The other two scales of the general system are zero: $R_2 = R_3 = 0$. The matrix of scales become proportional to the projector along the direction $(1, 0, 0)$. The second rotation $G'$ lost the angle around the last straight line direction, and therefore it is a function of two angles $\theta, \phi$, the angles which are obtained rotating from the original position of the tetrahedron to the position where the third coordinate coincides with the scaled coordinate along the straight line, namely
$$
\left(\begin{array}{cccc}
x_1 & x_2 & x_3 & x_4 \\
y_1 & y_2 & y_3 & y_4 \\
0 & 0 & 0 & 0
\end{array} \right) =
$$
\begin{equation}
 \left(\begin{array}{ccc}
\cos \psi & - \sin \psi & 0 \\
\sin \psi & \cos \psi & 0 \\
0 & 0 & 1
\end{array} \right) \left(\begin{array}{ccc}
R & 0 & 0 \\
0 & 0 & 0 \\
0 & 0 & 0
\end{array} \right) {\bf G}' {\bf E}\, ,
\end{equation}
where ${\bf G}'$ is parameterized by
\begin{equation}
{\bf G}' = \left(\begin{array}{ccc}
\cos \theta & \sin \theta \cos \phi & \sin \theta \sin \phi \\
- \sin \theta & \cos \theta \cos \phi & \cos \theta \sin \phi \\
0 & - \sin \phi & \cos \phi
\end{array} \right)\, .
\end{equation}
This is also expressed as
\begin{equation}
\left(\begin{array}{cccc}
x_1 & x_2 & x_3 & x_4 \\
y_1 & y_2 & y_3 & y_4 \\
0 & 0 & 0 & 0
\end{array} \right) = \left(\begin{array}{c}
R \cos \psi \\
R \sin \psi \\
0
\end{array} \right) \left(\begin{array}{ccc}
\cos \theta & \sin \theta \cos \phi & \sin \theta \sin \phi
\end{array} \right) {\bf E}\, .
\end{equation}

\begin{figure}
\centering
\scalebox{0.5}{\includegraphics{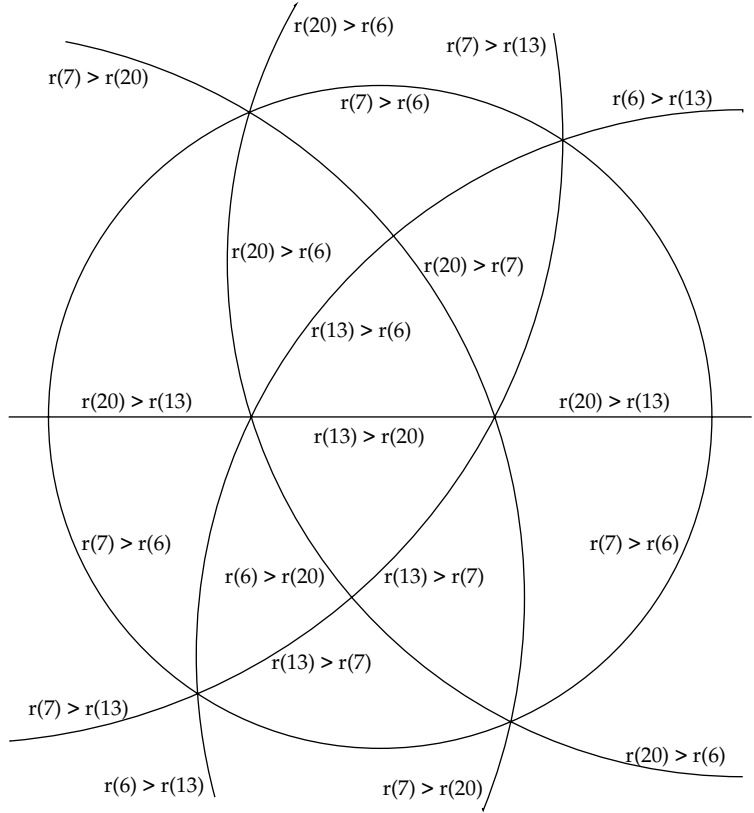}}

\caption{The set of the four-body collinear configurations in the central circle by the stereographic projection of the hemisphere of the two angles position of the orthocentric tetrahedron. The great circles represent the $r_i$ positions where two particles have the same coordinate. Each point on the hemisphere represent a different collinear configuration. The particular permutation is determined by the inequalities of the coordinates associated to two masses on both sides of the great circles. Each of the interior of the twelve spherical triangles represent the set of different collinear configurations with the same permutation order. The values of the masses are $m_1=20, m_2=13, m_3=7, m_4=6$. The inequalities between coordinates in figure are labeled with the numerical value of the corresponding masses. To simplify compilation the r's in figure are not italic fonts.}
\end{figure}

The number of independent coordinates for a collinear configuration is four. The coordinate $\psi$ determines the rotation of the line. The coordinate $R$ measure the expansion of the particles in the line. Its inertia moment is equal to $\mu R^2$. The geometric configuration is determined by the two angles $\theta$ and $\phi$
\begin{equation}
\left(\begin{array}{cccc}
r_1 & r_2 & r_3 & r_4
\end{array} \right) = \left(\begin{array}{ccc}
\cos \theta & \sin \theta \cos \phi & \sin \theta \sin \phi
\end{array} \right) {\bf E}\, ,
\end{equation}
equivalently
\begin{equation}
r_i = a_i \cos \theta + b_i \sin \theta \cos \phi + c_i \sin \theta \sin \phi \, .
\end{equation}
These are the coordinates along the collinear direction where the projection of the orthocentric tetrahedron is performed. They obey the normalization
\begin{equation}
\sum_{i = 1}^4 m_i r_i^2 = \mu \, , \label{norm}
\end{equation}
which is a consequence of equation (\ref{ort}).

The hemisphere of the two coordinates $\theta$ and $\phi$ is sectioned in twelve disjoint spherical triangles separated by the six main circles in the intersection of the sphere with the planes touching the orthocenter and orthogonal to the edges of the orthocentric tetrahedron. Each of these circles is the geometric set where two coordinates $r_i$ have the same value. On both sides of these circles the two corresponding coordinates obey a different inequality. Each spherical triangle is the set of points with a different permutation of the size of the four coordinates $r_i$. See figure 1 to illustrate these properties.

\section{Equations of Motion}
I assume for simplicity that the potential energy is given by the Newton potential (the gravitational constant is equal to 1)
\begin{equation}
V = - {1 \over R} \sum_{i<j}\frac{m_i m_j}{|r_i - r_j|}\, ,
\end{equation}
although our results may be generalized for any potential with a given power law of the relative distances between particles $R \, |r_i - r_j|$.

I also computed the kinetic energy as a function of the new coordinates, which is given by
\begin{equation}
T = {\mu \over 2} \left[ \dot{R}^2 + R^2 (\dot{\psi}^2 + \Omega_2^2 + \Omega_3^2)  \right] = {\mu \over 2} \left[ \dot{R}^2 + R^2 (\dot{\psi}^2 + \dot{\phi}^2 \sin^2 \theta + \dot{\theta}^2)  \right]\, ,
\end{equation}
where $\dot{\psi}$ is the angular velocity of the first rotation, and $(\Omega_1, \Omega_2, \Omega_3) = (\dot{\phi} \cos \theta, - \dot{\phi} \sin \theta, \dot{\theta})$ is the corresponding angular velocity vector of the second rotation.

The equations of motion follow from the Lagrange equations derived from the Lagrangian $T - V$ as presented in any standard text on Mechanics \cite{ll}, \cite{js}.

However, the $\psi$ coordinate related to the first rotation produces conservation of the angular momentum vector in the inertial system
\begin{equation}
\frac{\partial T}{\partial \dot{\psi}} = \mu R^2 \dot{\psi} = P_\psi \, , \label{mom}
\end{equation}
where $P_\psi$ is the numerical value of the conserved angular momentum.

The Lagrangian equation of motion for the scale coordinate is
\begin{equation}
\mu \frac{d^2}{d t^2} R + \mu R (\dot{\psi}^2 + \dot{\phi}^2 \sin^2 \theta + \dot{\theta}^2) = - \frac{\partial V}{\partial R} \, ,
\end{equation}

The Lagrangian equations of motion for the two coordinates associated with the second rotation are
\begin{equation}
\frac{d}{d t} (\mu R^2 \dot{\phi} \sin^2 \theta) = - \frac{\partial V}{\partial \phi} \label{fi}
\end{equation}
and
\begin{equation}
\frac{d}{d t} (\mu R^2 \dot{\theta}) = \mu R^2 \dot{\phi}^2 \sin \theta \cos \theta - \frac{\partial V}{\partial \theta} \, .\label{teta}
\end{equation}

There is one  more constant of motion, namely the total energy
\begin{equation}
E = T + V = {\mu \over 2} \left[ \dot{R}^2 + R^2 (\dot{\psi}^2 + \dot{\phi}^2 \sin^2 \theta + \dot{\theta}^2)  \right] + V \, .
\end{equation}

Writing the kinetic energy in terms of the angular momentum constant of motion instead of the $\dot{\psi}$ velocity lead us to
\begin{equation}
T = {\mu \over 2} \left[ \dot{R}^2 + R^2 (\dot{\phi}^2 \sin^2 \theta + \dot{\theta}^2) \right] +\frac{P_\psi^2}{2 \mu R^2}\, .
\end{equation}
The energy conservation is thus expressed as
\begin{equation}
E = {\mu \over 2} \left[ \dot{R}^2 + R^2 (\dot{\phi}^2 \sin^2 \theta + \dot{\theta}^2 ) \right] + \frac{P_\psi^2}{2 \mu R^2} + V \, , \label{ene}
\end{equation}
where $V$ represents the potential energy.

\section{Central configurations}

The collinear central configurations are characterized in our coordinates by constant values of the $\theta$ and $\phi$ angles. For these cases the angular velocity vector $\Omega$ is the null vector, and the conservation equations of moment and energy, (\ref{mom}) and (\ref{ene}) respectively, become
\begin{equation}
P_\psi = \mu R^2 \dot{\psi} \, .
\end{equation}
and
\begin{equation}
E = {\mu \over 2} \dot{R}^2 +\frac{P_\psi^2}{2 \mu R^2} + V \, .
\end{equation}
Note that the potential function $V$ becomes equal to a constant divided by coordinate $R$.

\begin{table}
\centering

Collinear central configurations with masses 20, 13, 7, 6.

\

\begin{tabular}{|r||r|}\hline
$\theta = $ 1.37525057217299 & $\theta =$ 0.983901787931397 \\
$\phi = $ 0.519159557815111 & $\phi =$ 5.11010135422999 \\ \hline
$r(20) = $ 0.366090733643358 & $r(13) = $ 0.486716343030589 \\
$r(13) = -$0.065366601796564 & $r(7) = $ 0.165320318621323 \\
$r(7) = -$0.373701028513627 & $r(20) = -$0.193830787117565 \\
$r(6) = -$0.642690274986074 & $r(6) = -$0.601323157899269 \\ \hline \hline
$\theta = $ 1.09192401664393 & $\theta =$ 0.520721995195208 \\
$\phi = $ 0.995617232736947 & $\phi =$ 4.55604682457794 \\ \hline
$r(20) = $ 0.371249801050636 & $r(7) = $ 0.599543803628696 \\
$r(7) = $ 0.012198842811027 & $r(13) = $ 0.246199851994839 \\
$r(13) = -$0.288938654946377 & $r(20) = -$0.189462920315934 \\
$r(6) = -$0.625697567731166 & $r(6) = -$0.601357715835848 \\ \hline \hline
$\theta = $ 1.4087509568619 & $\theta =$ 1.09821386579505 \\
$\phi = $ 1.43140526075336 & $\phi =$ 4.06952228046042 \\ \hline
$r(20) = $ 0.372886439009567 & $r(7) = $ 0.602862162854886 \\
$r(7) = $ 0.022933465266316 & $r(13) = $ 0.259974993861172 \\
$r(6) = -$0.200419706448564 & $r(6) = -$0.027293561159533 \\
$r(13) = -$0.493518830643398 & $r(20) = -$0.371797434661112 \\ \hline \hline
$\theta = $ 0.990022337434498 & $\theta =$ 1.38056997484206 \\
$\phi = $ 2.0613259406332 & $\phi =$ 3.65888040700007 \\ \hline
$r(7) = $ 0.580796256403095 & $r(7) = $ 0.619712718177074 \\
$r(20) = $ 0.166846782575173 & $r(6) = $ 0.35795101360832 \\
$r(6) = -$0.178777559500521 & $r(13) = $ 0.064367025252091 \\
$r(13) = -$0.486911083794 & $r(20) = -$0.366123321858331 \\ \hline \hline
$\theta = $ 1.37507165958022 & $\theta =$ 1.40879758950856 \\
$\phi = $ 2.58910007135665 & $\phi =$ 4.56889891614721 \\ \hline
$r(7) = $ 0.612335764627277 & $r(13) = $ 0.492293966825734 \\
$r(6) = $ 0.343103558437492 & $r(7) = $ 0.185550956125581 \\
$r(20) = -$0.012492053731027 & $r(6) = -$0.037738503416426 \\
$r(13) = -$0.468856202184258 & $r(20) = -$0.373612362055753 \\ \hline \hline
$\theta = $ 1.37010123641299 & $\theta =$ 0.517906500961236 \\
$\phi = $ 5.72909026677192 & $\phi =$ 1.65998590626889 \\ \hline
$r(13) = $ 0.4689161624149 & $r(7) = $ 0.58093101386497 \\
$r(20) = $ 0.011823000179912 & $r(20) = $ 0.162048815317523 \\
$r(7) = -$0.359730715863304 & $r(13) = -$0.275081849752453 \\
$r(6) = -$0.635709183991471 & $r(6) = -$0.621904892770559\\ \hline

\end{tabular}

\end{table}

These equations are identical to similar equations obtained for the Euler and Lagrange central configurations of the Three-Body problem \cite{bp}. They are formally the same as the equations for the conics in the Two-Body problem in terms of the radius $R$ and the true anomaly $\psi$.

\begin{figure}
\centering
\scalebox{0.5}{\includegraphics{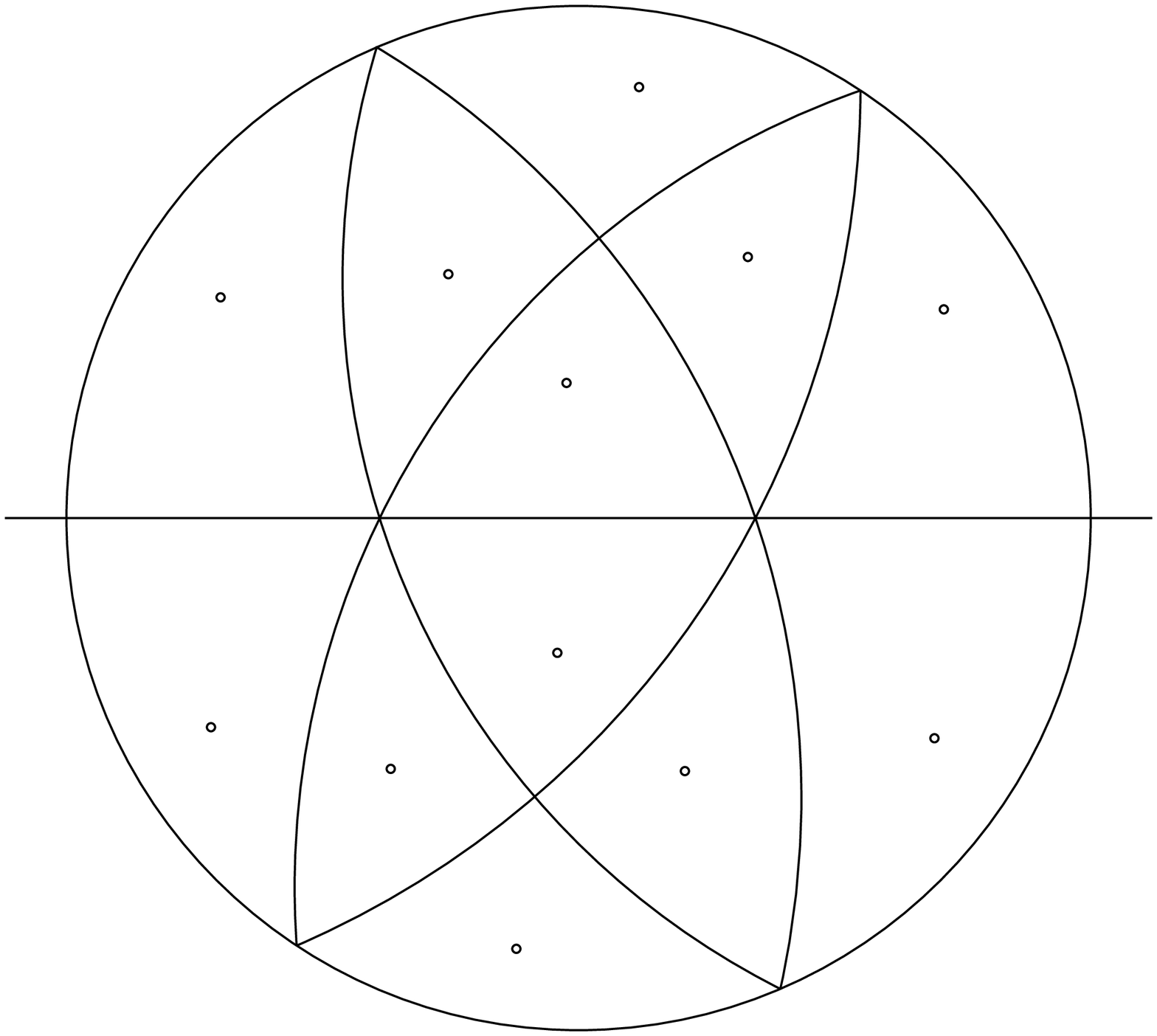}}

\caption{Stereographic projection of the hemisphere of the two angles motion of the orthocentric tetrahedron. The great circles are the positions where two particles have the same coordinate. Double intersection with a right angle of these circles are double-double collisions. Triple intersection is at triple collision. Each of the twelve spherical triangles have a right angle and two vertices of triple collision. The values of the masses were $m_1=20, m_2=13, m_3=7, m_4=6$. The centers of the small circles are at the positions where the central configurations were computed in the table.}
\end{figure}

\begin{figure}
\centering
\scalebox{0.5}{\includegraphics{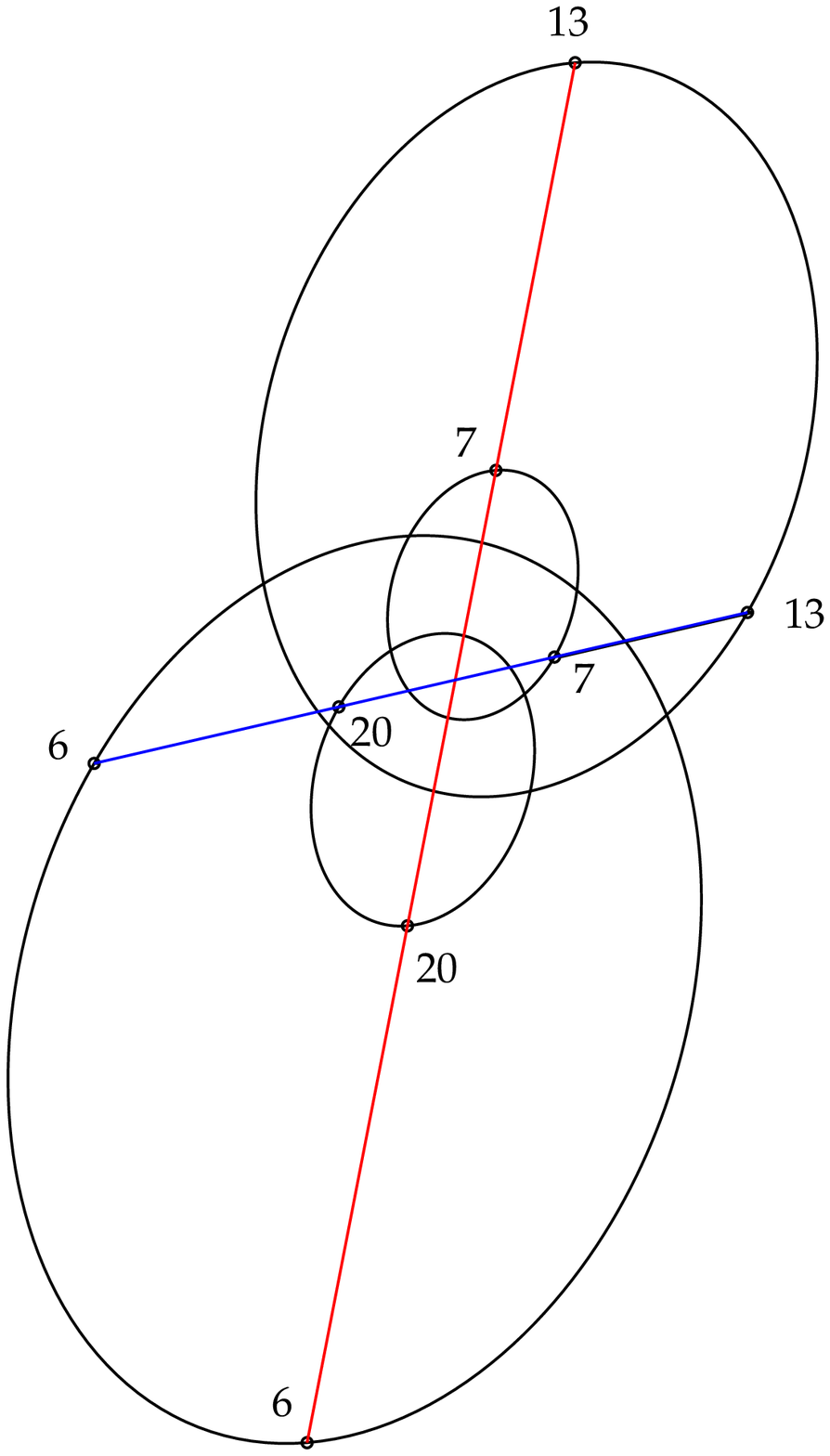}}

\caption{Two different times for the motion following a four-body simultaneous collinear central configurations. The values of the masses are $m_1=20, m_2=13, m_3=7, m_4=6$ with the permutation order 13, 7, 20, 6. The ellipses are labeled with the numerical value of the corresponding masses, eccentricity was chosen 0.7 to enhance the elliptic shape. The particular permutation from the table of central configurations is chosen disregarding the cases where one ellipse is very small, and the cases when two ellipses are very same equal, as is deduced from the values of the tabulated coordinates $r_i$. The center of mass is at the common focus, at the intersection of the two straight lines.}
\end{figure}

The constant values of the angles $\theta$ and $\phi$ referred to above are not arbitrary but they are determined by the two equations of motion (\ref{fi}) and (\ref{teta})
\begin{equation}
\frac{\partial V}{\partial \phi} = 0 \, , \quad \frac{\partial V}{\partial \theta} = 0 \, .\label{critical}
\end{equation}

These two equations are equivalent to the equations for central configurations that are presented generally as Moulton did \cite{mu}, written in terms of our notation as the four equations
\begin{equation}
\frac{\partial V}{\partial r_i} = \lambda m_i r_i \quad \mbox{($i$ = 1, 2, 3, 4)}\, ,\label{central}
\end{equation}
where $\lambda$ is the same parameter for the four equations. Using that the sum over the index $i$ is zero on both sides of this equation because the origin of coordinates is the center of mass, it follows that only three of them are independent. Canceling $\lambda$ one has only two independent equations. The two independent equations are obtained from equation (\ref{norm}) by taking the derivative of both sides with respect to $\theta$ and $\phi$\
\begin{equation}
\sum_{i = 1}^4 \frac{\partial r_i}{\partial \theta} m_i r_i = 0 \, , \quad \sum_{i = 1}^4 \frac{\partial r_i}{\partial \phi} m_i r_i = 0 \, .
\end{equation}
Substitution of equation (\ref{central}) in these equations lead to the equivalent conditions (\ref{critical}).

In order to compute four body collinear configurations one uses those conditions (\ref{critical}) by defining the two 4-vectors $s_i = (\partial r_i/ \partial \theta)$ and $\sin \theta \, t_i = (\partial r_i/ \partial \phi)$
\begin{equation}
s_i = - a_i \sin \theta + b_i \cos \theta \cos \phi + c_i \cos \theta \sin \phi \, , \quad t_i = - b_i \sin \phi + c_i \cos \phi \, .
\end{equation}

The two conditions to obtain the central configurations are therefore
\begin{equation}
\sum_{i = 1}^4  s_i \frac{\partial V}{\partial r_i} = 0 \, , \quad \sum_{i = 1}^4 t_i \frac{\partial V}{\partial r_i} = 0 \, .
\end{equation}
Which are two equations to determine the two angles $\theta$ and $\phi$.

Starting from an initial value for the two angles in the interior of the spherical triangle allowed by the order of masses in the line, we iterate looking for the root of one angle assuming the given value for the other alternatively in the two previous conditions and then one iterates until desired convergence.

Choosing the values of the masses as $m_1 = 20, m_2 = 13, m_3 = 7, m_4 = 6$, I have been able to compute the twelve different central configurations which are grouped in the table and represented his position in the hemisphere of angles $\theta, \phi$ by its stereographic projection in the figure 2.

The four particles moving in a central configuration is illustrated by the last figure. The four masses move in similar conics that in figure 3 were selected ellipses of eccentricity 0.7. The orbits of the ellipses comes from integration of equations (28) and (29) as
\begin{equation}
R = \frac{a (1- \epsilon^2)}{1 - \epsilon \cos(\psi - \psi_0)}\, ,
\end{equation}
where $\epsilon$ is the eccentricity and $a$ is the major semi-axis of the ellipse. The constant of integration $\psi_0$ measure the orientation of the orbits, and in figure 3 was chosen equal to $2 \pi/5$. The explicit coordinates for the four orbits were computed as function of $\psi$ from
\begin{equation}
\left( \begin{array}{rrrr}
x_1 & x_2 & x_3 & x_4 \\
y_1 & y_2 & y_3 & y_4
\end{array} \right) = \frac{a (1- \epsilon^2)}{1 - \epsilon \cos(\psi - \psi_0)} \left( \begin{array}{c}
\cos \psi \\
\sin \psi
\end{array} \right) (r_1 \; r_2 \; r_3 \; r_4)\, .
\end{equation}
The orbits are scaled from the ellipse $R, \psi$ by the four coordinates $r_i$ of the central configuration tabulated for the particular permutation 13, 7, 20, 6.

\end{document}